\documentclass[useAMS,usenatbib,usegraphicx]{mn2e}

\title[M15: Faint sources]
  {The X-ray source population of the globular cluster M15: 
   Chandra high resolution imaging}
\author[D.C. Hannikainen et al.]
  {D.C.~Hannikainen,$^{1,2}$, P.A.~Charles$^1$, L.~van Zyl$^{3,4}$,
  A.K.H. Kong$^5$, L. Homer$^6$ 
  \and P. Hakala$^2$, T. Naylor$^7$, and M.B. Davies$^8$ \\
  $^1$ Department of Physics and Astronomy, 
       University of Southampton, Southampton SO17 1BJ, UK\\
  $^2$ Observatory, PO Box 14, FIN-00014 University of Helsinki,
       Finland \\ 
  $^3$ Astrophysics Group, School of Physics, Keele University
       Staffordshire, ST5 5BG, UK \\
  $^4$ Department of Astrophysics, Oxford University,
       Keble Road, Oxford OX1 3RH, UK \\
  $^5$ Harvard-Smithsonian Center for Astrophysics, 60 Garden Street, 
       Cambridge, MA 02138, USA \\
  $^6$ Department of Astronomy, University of Washington, Box 351580, Seattle, WA 98195-1580, USA \\
  $^7$ School of Physics, University of Exeter, Stocker Road, Exeter
       EX4 4QL, UK \\
  $^8$ Lund Observatory, Box 43, SE-221 00 Lund, Sweden}
  
\def\LaTeX{L\kern-.36em\raise.3ex\hbox{a}\kern-.15em
    T\kern-.1667em\lower.7ex\hbox{E}\kern-.125emX}

\def\msun{${\rm M}_{\odot}$}
\def\rsun{${\rm R}_{\odot}$}
\def\lx{$L_{X}$}
\def\lo{$L_{opt}$}
\def\deg{$^{\circ}$}
\def\siml{\hbox{${_<\atop{\sim}}$}}
\def\simg{\hbox{${_>\atop{\sim}}$}}

\begin{document}

\label{firstpage}

\maketitle

\begin{abstract}
The globular cluster M15 was observed on three occasions with 
the High Resolution Camera on board {\it Chandra} in 2001 
in order to investigate the X-ray source population in the cluster 
centre.  
After subtraction of the two bright central sources, four 
faint sources were identified within 50 arcsec of the core.
One of these sources is probably the planetary nebula, K648, making 
this the first positive detection of X-rays from a planetary nebula 
inside a globular cluster.  
Another two are identified with UV variables (one previously known),
which we suggest are cataclysmic variables (CVs). 
The nature of the fourth source is more difficult to ascertain, and
we discuss whether it is possibly a quiescent soft X-ray transient
(qSXT) or also a CV.

\end{abstract}

\begin{keywords}
globular clusters: individual: M15 -- X-rays: binaries -- planetary
nebulae: individual: K648 -- stars: dwarf novae
\end{keywords}

\section{Introduction}

M15 is renowned for containing the first globular cluster 
Low-Mass X-ray Binary (LMXB; X2127+119) to be
optically identified (with AC211, see Auri\`ere, Le Fevre \& Terzan
1984 and Charles, Jones \& Naylor 1986), a counterpart that is still
one of the most optically luminous of all LMXBs.
AC211 is also the first globular cluster X-ray source
to have both an optical and X-ray
demonstration of its 17.1 hour orbital period via extended, partial
eclipses (see Ilovaisky et al. 1993 and 
Ioannou et al. 2002 and references therein).  Subsequent
studies of M15 show that it contains additional exotic, highly evolved
binaries. For example, there are at least 8 millisecond pulsars (MSP),
6 of them within 8 arcsec of the cluster core (Kulkarni \& Anderson
1996). Two of these MSPs have a negative $\dot{\rm P}$ (i.e. an
acceleration toward us as they move in the cluster potential) implying
a mass concentration of $\sim$ 4000
${\rm M}_{\odot}$ at the cluster centre (Phinney 1996). Phinney argues
that this mass {\it cannot} consist of low-mass stars and is either a
central black hole, or more likely a concentration of stellar
remnants: neutron stars and white dwarfs.

M15 is one of the most metal-poor globular clusters known ([Fe/H]
$\sim -$2.1) and has a low reddening factor (E$_{B-V} = 0.10 \pm
0.01$; Harris 1996).
A recent study has refined the distance to M15 to 9.98$\pm$0.47~kpc
(McNamara, Harrison \& Baumgardt 2004) which places it 5\% closer than
the previously quoted value of 10.4$\pm$0.8~kpc (Durrell \& Harris
1993).
McNamara et al. (2004) also provide a new estimate for the age and the
mass of the cluster which are 13.2$\pm$1.5~Gyr and (4.5$\pm$0.5)$\times
10^5$~\msun\, respectively.
M15's main-sequence turn-off point lies at $V \simeq 19.4, B-V \simeq 0.47$.

M15 was for a long time believed to be a classic example of a
central-surface-brightness-cusp globular cluster (Newell \& O'Neill
1978; Djorgovski \& King 1986; Peterson, Seitzer \& Cudworth 1989),
i.e. a cluster which has undergone core-collapse, with a small ($\sim$
0.06 pc), densely-crowded core containing many non-luminous remnants,
or even a single, massive black hole. However, more recent work shows
that M15's core, while compact, is larger than previously thought
(0.13 pc; Lauer et al. 1991), and the central velocity dispersion
lower ($\sim$ 10 km~s$^{-1}$; Gebhardt et al. 1992). As a result, it
is no longer possible to tell whether M15 has, in fact, undergone
core-collapse: a classical pre-collapse, thermal equilibrium cluster
model fits the data just as well as post-collapse models with a
re-expanding core (Grabhorn et al. 1992; De Marchi \& Paresce 1994).

Recently, van der Marel et al. (2002) and Gerssen et al. (2002, 2003),
using longslit spectra obtained with the Hubble Space Telescope (HST), have
shown that radial velocity measurements are consistent with a central
dark mass.  Models of the data which include a central black hole of
mass $\sim 2000$~\msun\, fit the data slightly better than models
without one. However, a central concentration of non-luminous, massive
stellar remnants fits the data just as well, as shown in the dynamical
simulations of Baumgardt et al. (2003).  Furthermore, the {\it Chandra}
X-ray images presented by White \& Angelini (2001) show no evidence
for any X-ray emission at the cluster centre, again indicating that it
is unlikely to harbour a central black hole (see also Ho, Terashima \& 
Okajima 2003).

One of the primary motivations for this project was to attempt to
solve one of X2127+119/AC211's biggest mysteries: the X-ray
and optical light curves, and its very low \lx/\lo, show unambiguously
that it is an accretion disc corona source (e.g. Ilovaisky et
al. 1993; Ioannou et al. 2002), in which the compact object is
obscured at all times by the accretion disc. But luminous type I X-ray
bursts have been observed from the system (e.g. Dotani et al. 1990),
requiring that a neutron star is visible (at least at certain times)! 
We inferred therefore that X2127+119 may actually consist of {\it two}
luminous X-ray binaries in M15's core, too close together to be
resolved by previous X-ray observations (Charles, Clarkson \& van Zyl
2002). This suggested exploiting {\it Chandra's} superb
$\sim$0.5-arcsec spatial resolution, and the result was the High
Resolution Camera (HRC-I) images presented here, in order to search
for additional sources close to AC211.

However, before the HRC-I observations were performed, White \&
Angelini (2001) serendipitously resolved X2127+119 into two bright
LMXBs with the zero-order ACIS-S/HETG image. M15~X-2 was found to be
associated with a $U\sim19$ star located 2.7 arcsec from AC211 and 3.3
arcsec NW of the cluster core. While M15~X-2 is $\sim$4 mag fainter
than AC211, it has an X-ray count rate $\sim$2.5 times
greater. Nothing is known yet about the nature of M15~X-2, other than
that -- if we assume it to be responsible for the type I X-ray bursts
(and thereby resolving the controversy of AC211's X-ray properties) --
the compact object is a neutron star. This then leaves the nature of
the object in AC211 unconstrained.

High resolution X-ray imaging is now revealing a wide range
of low-luminosity sources in globular clusters (see e.g. Heinke et
al. 2003a; Hakala et al. 1997).
The advent of {\it Chandra} has made it possible to investigate these sources
in earnest. Some of the first results, (e.g. Pooley et al. 2002a,b),
show detections of CVs in the globular clusters NGC~6752 and
NGC~6440. Dozens of low-luminosity X-ray sources have also been
discovered in $\omega$ Cen and NGC~6397 (see Verbunt \& Lewin 2004 
for a recent review), and $\sim$300 in 47 Tuc (Heinke et al. 2004).

These studies are narrowing the gap
between the predicted and observed CV population densities, but have
been primarily restricted to the cases of those clusters that do {\it
  not} contain central luminous steady LMXBs.
However, one should note that a number of studies has emerged recently using 
 {\it Chandra} to look for faint sources in globular clusters which do
 contain bright LMXBs, e.g. Heinke, Edmonds \& Grindlay (2001), Heinke
 et al. (2003b), Wijnands et al. (2002), Homer et al. (2002).

M15 is a good cluster to hunt for CVs and LMXBs because of its
large mass and high compactness: its very dense core should be an 
ideal breeding ground for CVs and LMXBs (as indicated already by the 
MSP population). 
In addition, its very low reddening will not hinder the UV/optical
identification of any new sources that are found.

What are the chances of optically detecting a dwarf nova (DN) 
in outburst in M15 during a single HST visit? Di Stefano \& Rappaport (1994)
predicted a population of $\sim$ 200 CVs in 47 Tuc, of which $\sim$ 50
are DNe. We would expect M15 to have a much higher population of CVs
because of its greater core mass and central condensation than 47 Tuc.
Nevertheless, assuming that M15 has say 50 DNe, with a mean outburst
recurrence time of 50 days and outburst length of 10 days, then at any
given time, one would expect to see $\sim$10 DNe in outburst.

Although DNe are brighter in the optical during outburst than
in quiescence, their behaviour in the X-ray is harder to predict.
A good summary of the current view of the origin of the X-rays is
given by Verbunt, Wheatley \& Mattei (1999).
For our purposes the crucial point is that in outburst the 2.5--10~keV
flux from the boundary layer largely disappears, and is replaced by a
very soft extreme ultraviolet (EUV) component.
The disappearance of the hard component should drive the X-ray flux
down in outburst, but the EUV component is so luminous that if its
hard energy tail reaches the soft X-ray range, it will dominate the 
flux there, and lead to a brightening in that range.
Thus whether the DN brightens or fades depends crucially on the 
X-ray energy range in question, and to a lesser extent the reddening.
Given that we know at least one DN (U Gem; Swank et al. 1978) whose
flux increases even in the 2--10~keV range, some DNe will be brighter in
outburst than quiescence in HRC-I, but it is not clear this is true for
all of them.

From an observational point of view, we are particularly interested in
the DNe subclass of CVs because of their frequent outbursts making
them easier to detect and classify as definite CVs. Outburst
recurrence intervals for DNe range typically from 14 -- 120 days or
longer, and last $\sim$10 days (see e.g. Warner 1995).

Hence, we use our {\it Chandra} data here to investigate these fainter
sources, in an attempt to reveal M15's menagerie of CVs, quiescent
low-mass X-ray binaries, (chromospherically) active binaries 
and other exotica for the
first time. This work is still in progress: detecting the faint X-ray
sources in the core, which is dominated by the flux from AC211 and
M15~X-2, has been -- and continues to be -- a considerable challenge as we
will discuss below.

In Section~\ref{chandra} we present our {\it Chandra} HRC-I observations,
with particular emphasis on how to remove the two luminous sources so
as to reveal the faint population within. In Section 3, we
use archival HST imaging to search for optical counterparts to these
faint X-ray sources and discuss the implications of these results for
the nature of this faint source population.

\section{{\it Chandra} HRC Imaging}\label{chandra}

\subsection{Observations}

We obtained three separate observations of M15 with {\it Chandra's}
  HRC-I (Murray et al. 1997) of approximately 10~ksec each over a six
  week period in 2001~Jul--Aug. 
The HRC-I has a $30^{\prime}\times30^{\prime}$ field of view. 
Table~\ref{tab-obs} summarizes the details of the observations. 
All data were reduced and analyzed using {\sc CIAO} versions 2.0 to 2.3.

\begin{figure}
\includegraphics[width=80mm]{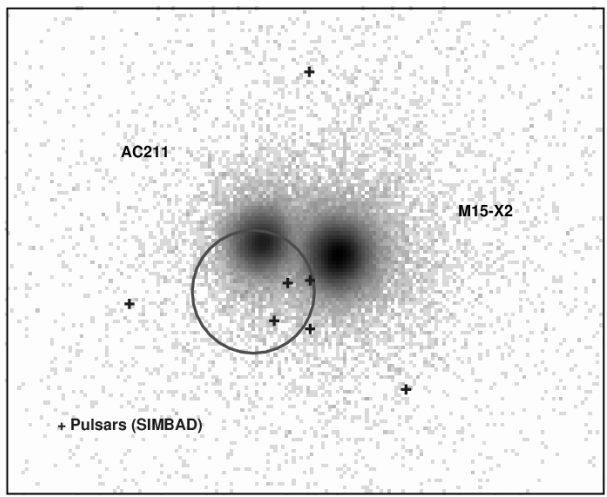}
\caption{Our summed HRC-I image of the centre of M15, showing the two 
 bright sources, AC211 and M15~X-2, and the positions of 7 of the
8 known millisecond pulsars (from {\sc SIMBAD}). The circle represents the
 location and extent of the cluster core (which has a radius of 2.2
 arcsecs, see Lauer et al. 1991).  }
\label{fig-m15}
\end{figure}

Figure~\ref{fig-m15} shows our summed HRC-I image of the central region
  of M15. 
The crosses show the positions of 7 of the 8 known MSP's, while
  the circle represents the $2^{\prime\prime}.2$ core radius centred on 
  the position of the cluster centre ($\alpha=21^h29^m58.^s335, 
  \delta=12^{\circ}10^{\prime}0.^{\prime\prime}89$, J2000; van der Marel 
  et al. 2002). 
The {\it Chandra} and HST images (Sec. 3) were realigned on the basis of the
  AC211 positional match.

\begin{table}
\caption{{\it Chandra} observation log}
\label{tab-obs}
\begin{tabular}{cccc} 
Observation number & Date        & UT$_{\rm start}$    & UT$_{\rm stop}$ \\\hline
1903               & 2001 Jul 13 & 07:07:43           & 10:04:42 \\ 
2412               & 2001 Aug 3  & 16:24:35           & 19:23:48 \\
2413               & 2001 Aug 22 & 04:57:07           & 08:23:19 \\ \hline
\end{tabular}
\end{table}

\begin{figure*}
\includegraphics[width=160mm]{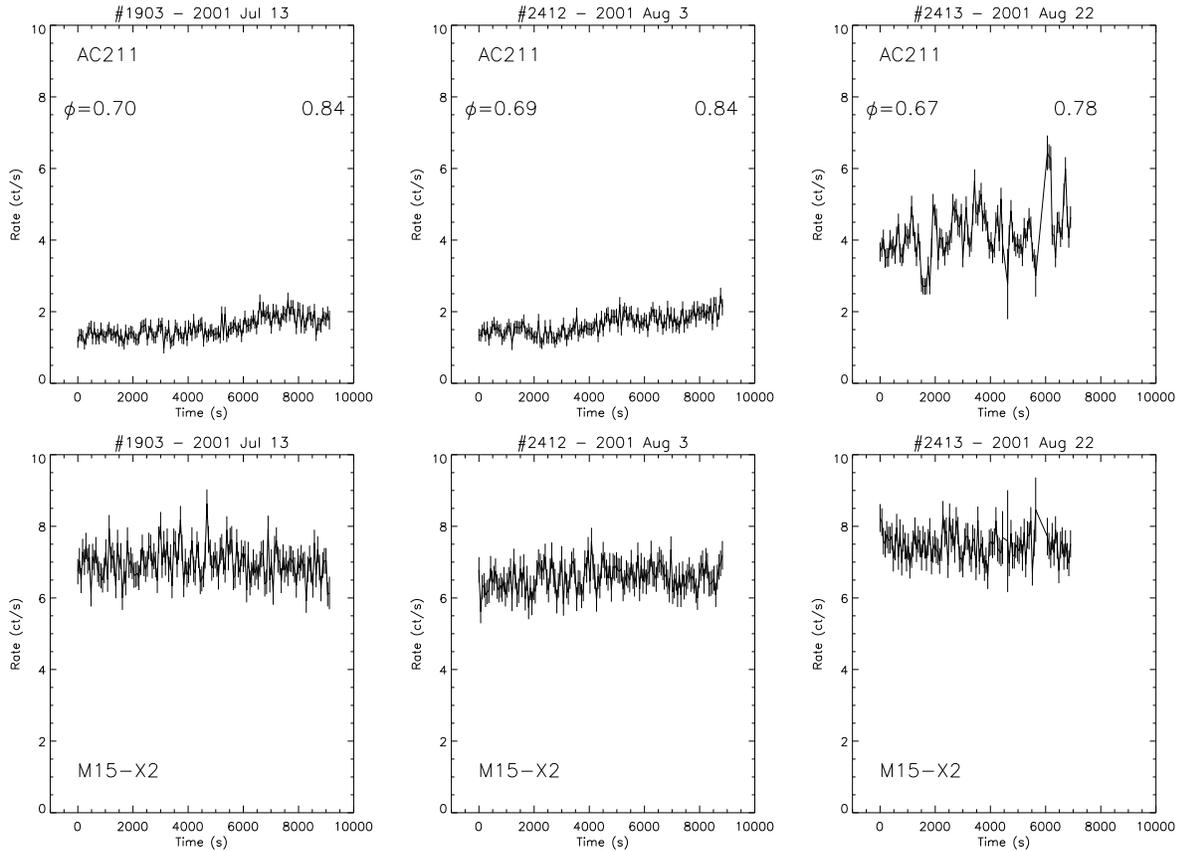}
\caption{The light curves of AC211(top) and M15~X-2 (bottom) from the
  three observation dates. The binary phase of AC211 is given on the
  plots at the beginning and end times. 
}
\label{fig-lc}
\end{figure*}

Figure~\ref{fig-lc} shows the light curves obtained for AC211 and M15
  X-2 during the three observations. 
The phase of each observation,
  obtained using the ephemeris in Ioannou et al. (2002), 
  is given in each plot.  
They show that the
  observations were undertaken just prior to the eclipse in AC211.  
It is clearly seen that M15~X-2 is a steady source with an average count
  rate of 6.8 ct/s, while AC211 enters a flaring state during the last
  observation on 2001 Aug 22.

Table~\ref{tab-fl} shows the average count rates and the fluxes
  derived for both sources using the spectral parameters from White \&
  Angelini (2001).  
For M15~X-2 we fixed the hydrogen column density to the expected value
  of M15, i.e. $N_H=6.7\times10^{20}$~cm$^{-2}$, and used a power law
  with an index of 1.89 to derive the 0.5--7.0~keV luminosity.  
In the case of AC211, we used the partial covering model 
  with a power law photon
  index of 2.0, a covering fraction of 0.92 and an intrinsic 
  absorption of $2.05\times10^{22}\,{\rm cm}^{-2}$.
Again, the interstellar absorption was set to that of M15.
Assuming a distance of 9.98~kpc to M15, these
  convert to an average luminosity of $L_x=1.4\times10^{36}~{\rm
  erg}~{\rm s}^{-1}$ for M15~X-2 (consistent with that found by White \&
  Angelini, although note that they used the previous value for the
  distance to the cluster of 10.3~kpc) 
  while AC211 varies between $L_x=7.9\times10^{35}~{\rm
  erg}~{\rm s}^{-1}$ in the first observation to
  $L_x=2.2\times10^{36}~{\rm erg}~{\rm s}^{-1}$ in the last observation.

The luminosity of M15~X-2 is consistent with that of White \& Angelini
(2001), as are the luminosities for AC211 in observations 1903 and
2412. However, the luminosity of AC211 in observation 2413 is significantly
higher than in the other two observations, and indicates a brightening
of the source at that time. 

Table~\ref{tab-fl} also shows the count rate and X-ray luminosities
for source D, most likely identified with the planetary nebula (PN)
K648, as we discuss in more detail in Section~2.2. 

\begin{table*}
\caption{Count rates and luminosities}
\label{tab-fl}
\begin{center}
\begin{tabular}{ccccccc}\hline
Obs. no.  & \multicolumn{2}{c}{1903} & \multicolumn{2}{c}{2412} & \multicolumn{2}{c}{2413} \\\hline
  & Count rate  & L$_{\rm X}$ (0.5--7 keV)  & Count rate  &  L$_{\rm X}$ (0.5--7 keV)  & Count rate  & L$_{\rm X}$ (0.5--7 keV)  \\
  & (s$^{-1}$)  & (erg s$^{-1}$) & (s$^{-1}$)&  (erg s$^{-1}$) &(s$^{-1}$) & (erg s$^{-1}$) \\\hline\hline
AC211    & 1.55$\pm$0.18 & $7.9\times10^{35}$ & 1.65$\pm$0.19 & $8.4\times10^{35}$ & 3.87$\pm$0.27 & $2.2\times10^{36}$ \\
M15~X-2  & 6.97$\pm0.36$ & $1.4\times10^{36}$ & 6.57$\pm0.35$ & $1.3\times10^{36}$ & 6.95$\pm0.36$ & $1.5\times10^{36}$ \\
K648  & $0.26\pm0.18\times10^{-3}$ & $2.6-6.1\times10^{31}$$^{\star}$
  &$<0.06\times10^{-3}$ & $<0.6-1.4\times10^{31}$$^{\star}$ & $1.03\pm0.42\times10^{-3}$
  &$1.0-2.4\times10^{32}$$^{\star}$ \\ \hline \hline
\end{tabular}
\end{center}
\hspace{-140mm}{\footnotesize $^{\star}$ See Table~\ref{tab-xraysources} and Section 2.2.} 
\end{table*}

\subsection{Searching for faint sources}

The main problem in detecting faint sources in M15 is that, unlike 
for instance the
clusters studied by Pooley et al. (2002a,b) in which there are no
X-ray bright sources, M15 houses not just one but {\em two}
bright central sources. Any faint sources lurking near the centre of
the cluster -- which is where we would expect to find them -- will be
contaminated by the wings of the AC211 and M15~X-2 point spread functions
(PSF).  Thus to have any hope of detecting faint sources, the two
bright central sources must first be subtracted. We did
this by constructing a PSF for each of the two sources using the {\it
Chandra} PSF libraries. To increase the signal-to-noise of any
possible faint sources, we merged the three observations.  We then
used the faint source algorithm {\sc wavdetect} on the PSF-subtracted
merged image.

\begin{figure}
\includegraphics[width=80mm]{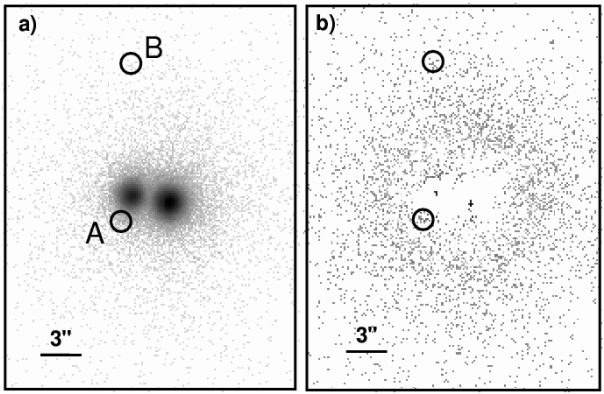}
\caption{{\bf (a)} The merged image of all three observations
  of the centre of M15, showing the two bright LMXBs.  
{\bf (b)} The image after PSF-subtraction of these sources; 
the two faint sources detected by the source detection algorithm
{\sc wavdetect} in this image are overlaid in both panels.}  
\label{fig-psf}
\end{figure}

\begin{figure}
\includegraphics[width=80mm]{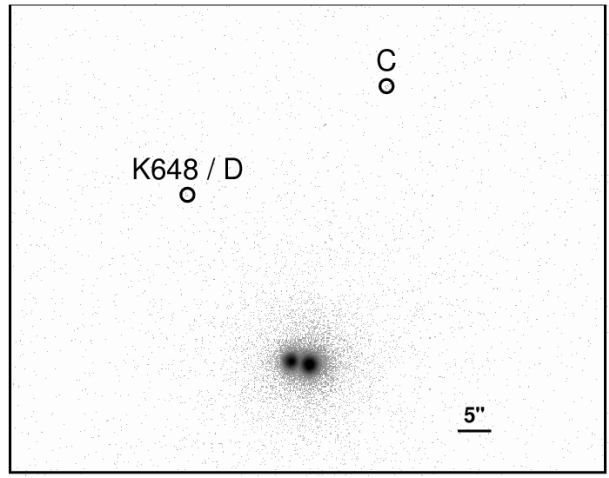}
\caption{{\it Chandra} HRC-I image showing the locations of a
possible dwarf nova (source C) and the planetary nebula, K648 (source D),
detected using {\sc wavdetect}, both well within the half-mass radius. 
The image is approximately 90$^{\prime\prime}$ (width) by 
70$^{\prime\prime}$ (height). The bar shows 5$^{\prime\prime}$.
}
\label{fig-dn+pn}
\end{figure}

Figure~\ref{fig-psf} shows both the merged image (a) and the resulting
PSF-subtracted image (b) of
the merged observations. They both show the positions of
potential sources (A and B) found after {\sc wavdetect} was run on the
PSF-subtracted image. As can be seen immediately from these images,
the software and understanding of the {\it Chandra} PSF is not yet
able to smoothly subtract bright sources from images without leaving
large-scale (but smooth) artefacts such as the central ``hole''.  
Nevertheless, it
does allow a more sensitive search for fainter sources to be
undertaken on the subtracted image, and we can gauge the effectiveness
of this analysis to search for potential counterparts 
by using archival Hubble Space Telescope images. In
addition to the sources found close to AC211 and M15~X-2, {\sc wavdetect} also
found two sources (C and D) at larger radii, one being the planetary
nebula K648, shown in Figure~\ref{fig-dn+pn}.

Table~\ref{tab-sources} shows the coordinates and {\it U}-magnitudes 
for the four sources found in and near the centre of M15.
Table~\ref{tab-xraysources} shows the X-ray luminosities
for these four sources estimated using typical spectral parameters,
based on our candidate IDs. 
For sources A and C (DNe) we used a Bremsstrahlung 
  temperature of 4~keV. 
The merged dataset was used to extract the
  parameters for source A. 
For source B (qSXT/DN?) we applied several models.
We used the parameters for Aql X-1 in {\it quiescence} 
(taken from Rutledge et al. 2001 using {\it Chandra} data). 
Two estimates were derived, one for a power law with photon 
  index 4.1 and one for a blackbody temperature of 0.33~keV.
In addition, we also used a Bremsstrahlung temperature of 4~keV. 
Finally, for source D (PN) we applied two models: we used
  parameters from Guerrero et al. (2001), i.e. a Raymond-Smith model with 
  temperature 0.64~keV, and -- in the case that the emission might
  arise from a compact region (as discussed in Section~3.3) -- we 
  applied a thermal Bremsstrahlung temperature of 4~keV. 
We discuss the nature of these sources 
  and compare them to HST data in the next section.

\begin{table*}
\caption{Optical counterparts of faint {\it Chandra} sources in M15}
\label{tab-sources}
\begin{center}
\begin{tabular}{cccccc} \hline
ID & Object&   Location$^{\star}$ &  Coordinates (J2000) &  {\it U}-magnitude &  Variable? \\ 
   &    &          &               RA DEC   & & \\ \hline 
A & DN    & 0.3'' NE & $21^h29^m58.37^s +12^{\circ}10^{\prime}00.8^{\prime\prime}$ & 17.6$^{\dagger}$ & Yes \\
B & qSXT/DN? & 11.'' N  & $21^h29^m58.32^s +12^{\circ}10^{\prime}12.4^{\prime\prime}$ &  21.8 & Possibly \\ 
C & DN    & 46.'' NE & $21^h29^m57.34^s +12^{\circ}10^{\prime}43.7^{\prime\prime}$ &  20.2$^{\dagger}$ & Yes \\
D & K648  & 31.'' NW & $21^h29^m59.38^s +12^{\circ}10^{\prime}27.5^{\prime\prime}$ & 13.5$^{\ddagger}$ &Yes (X-rays) \\\hline \hline
\end{tabular}
\end{center}
\hspace{-70mm}{\footnotesize $^{\star}$With respect to the cluster centre.}\\
\hspace{-40mm}{\footnotesize $^{\dagger}$Magnitude in outburst (object not visible in quiescence).}\\
\hspace{-70mm}{\footnotesize $^{\ddagger}$from Alves, Bond \& Livio (2000).}
\end{table*}

\begin{table*}
\caption{X-ray luminosities of faint {\it Chandra} sources in M15
  assuming a distance of 9.98~kpc}
\label{tab-xraysources}
\begin{center}
\begin{tabular}{cccccc} \hline
   \multicolumn{6}{c}{X-ray luminosities (erg s$^{-1}$ in the
   0.5--7~keV range)} \\ \hline
   & & & \multicolumn{3}{c}{Obs \#} \\
Source  &       &   Model  &                 1903   &   2412   & 2413 \\ \hline \hline
 A      &  DN   & kT$_{\rm Bremss}$=4 keV &         & $3.3\times10^{32}$$^{\dagger}$ & \\
 B      & qSXT? & $\alpha=4.1$  & $2.34\times10^{31}$ & $7.30\times10^{31}$ & $1.15\times10^{32}$ \\
        &       & kT$_{\rm BB}$=0.33 keV & $4.28\times10^{31}$ & $1.34\times10^{32}$ & $2.10\times10^{32}$ \\
        &  DN?  & kT$_{\rm Bremss}$=4 keV & $7.46\times10^{31}$ & $2.33\times10^{32}$ & $3.66\times10^{32}$ \\
 C      &  DN   & kT$_{\rm Bremss}$=4 keV & $8.65\times10^{31}$ & $1.94\times10^{32}$ & $1.66\times10^{32}$ \\
 D      & K648  & kT$_{\rm RS}$=0.64 keV & $2.6\times10^{31}$ & $<0.6\times10^{31}$ &
   $1.0\times10^{32}$ \\ 
        &       & kT$_{\rm Bremss}$=4 keV & $6.1\times10^{31}$ &
   $<1.4\times10^{31}$ & $2.4\times10^{32}$ \\ \hline
\end{tabular}
\end{center}
\hspace{-70mm}{\footnotesize $^{\dagger}$ The merged dataset was used
  to derive the luminosity of source A.}
\end{table*}

To estimate whether any of the four faint sources may actually be
  background sources, we use the log N -- log S relations of Giacconi et al.
  (2001).  
Using the count rate of our faintest source to set a flux
  detection limit (and assuming that it is in fact a typical
  background object with $\Gamma=1.5$), we obtain 
  $F_{lim} \simeq 2$ (0.5-2 keV) and 4.5 (2-10 keV) $\times10^{-15}
  {\rm erg}~{\rm s}^{-1}~{\rm cm}^{-2}$,
  and hence we expect $\simeq 0.5$ discrete background 
  sources within the area enclosed by the half-mass radius
  (1.06$^{\prime}$) of M15.  
This provides an upper limit since, even with our PSF subtraction,
  much of the central region of M15 is not probed
  to this depth. 
We conclude that at most one source could be a background
  object, but given our optical counterpart IDs (see next section) we would
  suggest that they are all in fact associated with M15.

\section{HST Imaging and the Nature of the Low $L_X$ Sources}

M15 has been a regular target for HST. Studies of M15's stellar
population have been carried out, for example, by Ferraro \& Paresce
(1993), De Marchi \& Paresce (1994; 1996) Guhathakurta et al. (1996)
and Sosin \& King (1997), and an intensive photometric study of M15's
PN, K648, was undertaken by Bianchi et al. (1995; 2001) and Alves,
Bond \& Livio (2000). The HST archives therefore contain many images
of M15 obtained at different epochs.

At the distance to M15, DNe would be expected to
have quiescent apparent magnitudes of 21 to 24, with outburst
magnitudes ranging from 14 to 21. Magnetic CVs would have magnitudes
similar to quiescent DNe (see Warner 1995), while the high mass
transfer rate CVs (the nova-likes and nova remnants) would be 2--3
magnitudes brighter. X-ray transients would be expected to have
quiescent apparent magnitudes similar to those of DNe, and outburst
magnitudes ranging from 13 to 18. The deepest HST archival images of
M15 go down to 22--23 mag. Therefore, most quiescent CVs and X-ray
transients would be too faint to detect in such images.

\subsection{Probable Dwarf Novae}

\begin{figure}
\includegraphics[width=80mm,angle=0]{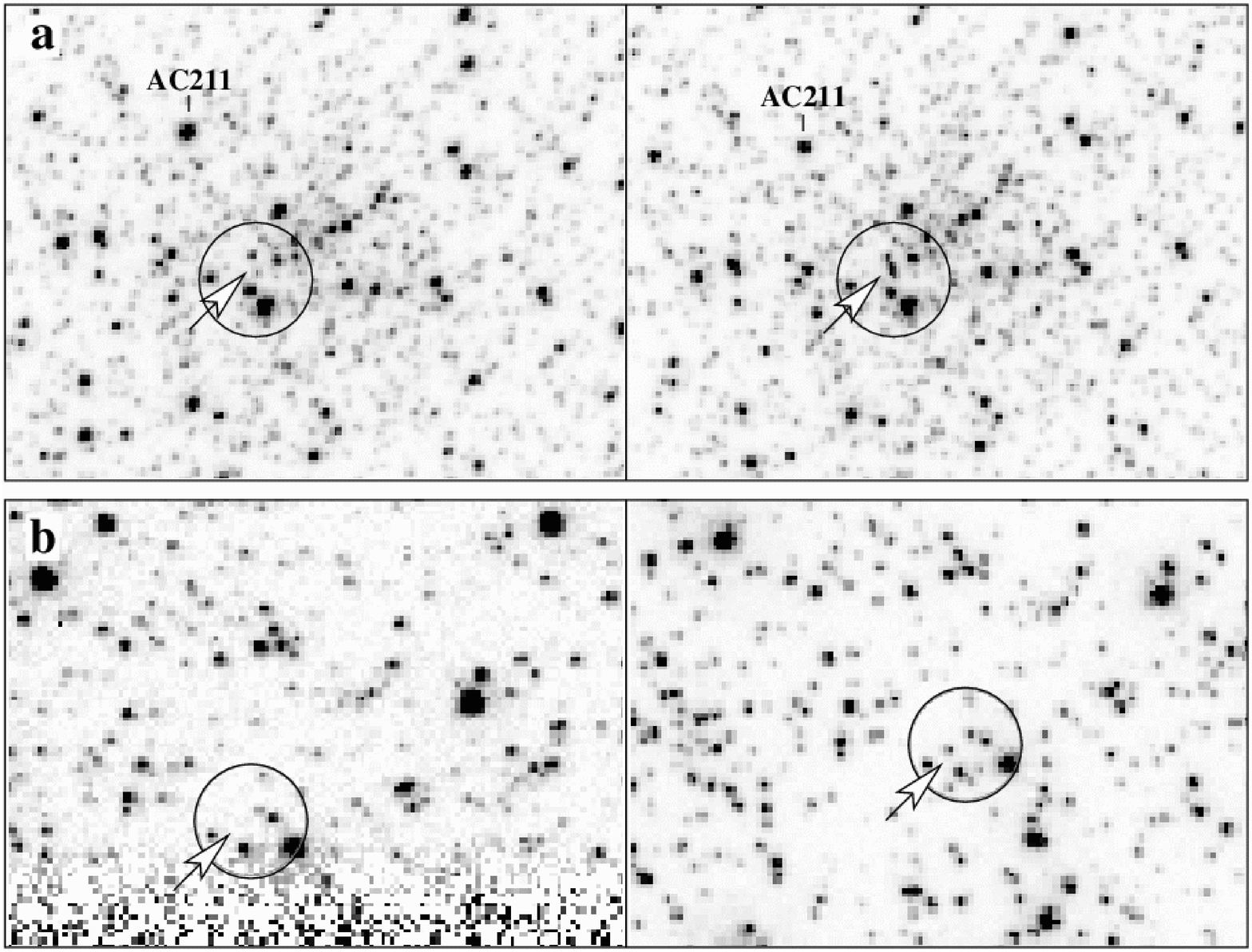}
\caption{(a) Probable dwarf nova (source A) in the core of M15, found
using image-subtraction, reported in Charles, Clarkson \& 
van Zyl (2002), found here to be associated with a {\it Chandra}
X-ray source. (b) Probable dwarf nova (source C) associated with a 
{\it Chandra} X-ray source, located $46^{\prime\prime}$ NE of
the cluster core. The error circles are the {\it Chandra}
0.25~arcsec error circle.
}
\label{fig-DNs}
\end{figure}

We looked for variables in M15's core by aligning and subtracting
images from different epochs, a technique which highlights stars which
vary between the images. In addition to detecting two
previously-identified RR Lyrae stars (Ferraro \& Paresce 1993), we
made a first detection of a probable DN in outburst
(Fig.~\ref{fig-DNs}a), located just 0.3 arcsec from the cluster core
(the optical discovery was reported in Charles, Clarkson \& van Zyl
2002). The object appears in WFPC2 {\it U}-band images taken in October 1994
with  {\it U}=17.60, but is undetectable in images obtained at
other epochs. Therefore its variability range is $>$5 magnitudes (the
deepest available images have a limiting {\it U}$\sim$22.5). This object is
revealed here to coincide with one of our
faint {\it Chandra} X-ray sources (source A); its luminosity of
$3.3 \times 10^{32}$~erg~s$^{-1}$ is consistent with the brighter CVs
identified in other globular clusters. 
Outbursting DNe often start off hard, but this quickly disappears to
be replaced by a much softer component (see e.g. the light curves
of SS Cyg in outburst in Kuulkers et al. 2003).
However, we have no way of knowing
at what stage in its activity cycle it has been observed.

If this object is indeed a DN or an X-ray transient, it is
statistically much more likely to be the former than the latter. DNe
outburst recurrence times are weeks to months, while X-ray transients
usually recur on much longer timescales, typically tens of years. The
chances of catching an X-ray transient in outburst on the few
occasions on which the HST has observed M15 over the last decade are
therefore extremely small.  The likelihood that we have a DN rather
than an X-ray transient depends also on the relative numbers of these
two kinds of object in the core of M15: 
we expect CVs to be more numerous than X-ray binaries.  
However, we do note that the (neutron star) transients in globular
clusters appear to have shorter recurrence times between outbursts
than those in the field (e.g. NGC~6440, Pooley et al. 2002b).  
With all of those in clusters having been 
identified as neutron star systems (due to the presence of type I 
X-ray bursts), they represent a different population from the soft 
X-ray transients overall, of which only $\sim$25\% contain neutron
stars, the rest being black-hole candidates (see e.g. Charles 2001).  
And since the well-known neutron star SXT Aql X-1 outbursts once or
twice per year (e.g. Simon 2002), 
it is still possible that our source A might not
turn out to be a DN.

The second probable DN was found in the error circle of a faint
{\it Chandra} X-ray source 46 arcsec NE of the cluster core (source C in
Figure~\ref{fig-dn+pn}, well within the cluster half-mass radius). It
appears in outburst with {\it U}=20.24 in images obtained in October 1994
(Fig.~5b), but is completely absent in images at other epochs. Again, 
as we have no constraint on the light curve there is no
way of telling whether we caught it at the peak of its outburst, but
this magnitude is consistent with the DN outburst range (14 to 21)
expected for M15. Of course, it is also possible that this object is
an X-ray transient, but using the same arguments as above, we
believe it is more likely to be a DN.

Tuairisg et al. (2003) conducted a ground-based search in 1997 for 
outbursting DNe in M15 but failed to find any. 
If these sources (A and C) do turn out to be outbursting DNe, then
these are amongst the few identified thus far in globular
clusters -- only a handful of erupting dwarf novae in globular
clusters have been confirmed (or suggested as possible outbursting DNe) 
to date (e.g. M22, Anderson, Cool \& King 2003; M5, Kaluzny et al. 1999; 
47~Tuc, Edmonds et al. 2003).

\subsection{Possible qSXT or DN?}

 \begin{figure}
\includegraphics[width=80mm,angle=0]{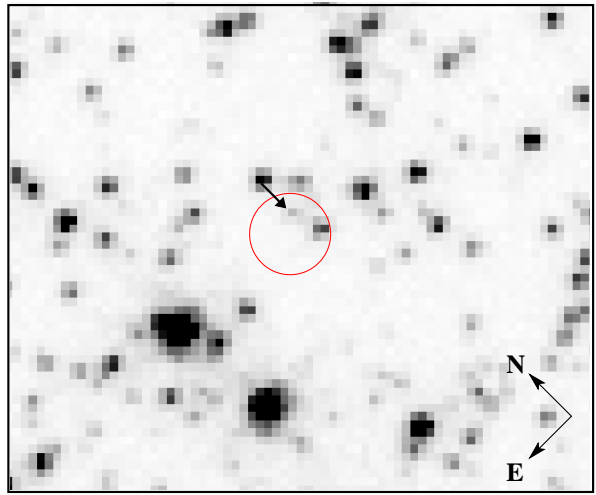}
\caption{Possible qSXT? The faint blue star indicated appears to be
  variable. The error circle is the {\it Chandra} 0.25
  arcsec error circle. 
}
\label{fig-maybeqSXT}
\end{figure}

The two sources discussed above are designated `probable' DNe, because
 they were obviously variable in the HST images and showed possible
 outbursts. The object discussed in this section is harder to
 classify, but is the only blue star near the corresponding
 {\it Chandra} X-ray source (Fig.~\ref{fig-maybeqSXT}). 
The source in question is source B in Fig.~\ref{fig-psf}
and the possible counterpart is listed in Table~3.
 As explained in Section~2.2, we calculated its X-ray luminosity based on
 several models -- two for a possible qSXT scenario and one for a
 possible DN scenario (see Table~4).

In all three cases, the X-ray source has a luminosity of a few
$\times10^{31}$ -- $10^{32}$~erg~s$^{-1}$. 
The candidate optical counterpart has {\it U}=21.8, {\it V}=22.54, 
and appears to show some variability ($\Delta V\sim0.04$),
although this is uncertain -- the object is extremely
faint and the apparent variability may simply be noise
(van Zyl 2002).
Its X-ray luminosity is more consistent with that of a CV, although 
it is also comparable to the faint end of the distribution of 
qSXTs in GCs (Heinke et al. 2003a).
With $U-V=-0.74$, its optical colours  are again consistent with 
those of the DNe SS Cyg and VW
Hyi in quiescence (Bailey 1980), similar to that of the NS SXT Cen X-4
in quiescence (McClintock \& Remillard 2000), but much bluer 
than the optically identified qSXT (X5) in 47 Tuc (Edmonds et al. 2002).
We therefore suggest that source B is more likely another CV 
rather than a qSXT.

\subsection{The Planetary Nebula K648}

The fourth faint X-ray source we detected has a position consistent 
  (within the {\it Chandra} uncertainties) with one of the rare globular cluster PNe,
  K648  (Alves, Bond \& Livio 2000; hereafter ABL). 
The source was seen to vary by a factor greater than three 
  during the {\it Chandra} observations, being observed at 
  $L_X=0.26$ and $1.0\times10^{32}$erg~s$^{-1}$ in two (\#1903,
  \#2413) but undetected in the third (\#2412). 
We can reject that the source is constant at the $>$98\% level.  
However, its luminosities are within the range 3$\times$10$^{29}$ 
  to 3$\times$10$^{32}$ erg~s$^{-1}$ in the
  0.4--1.7~keV band found for 13 X-ray-active PNe in the galactic
  plane using the ROSAT archive by Guerrero, Chu \& Gruendl (2000), 
  and for the well-resolved, extended X-ray emission detected with 
  {\it Chandra} from the PNe BD+30\deg3639 (Kastner et al. 2000) and
  NGC~7027 (Kastner, Vrtilek \& Soker 2001). 
Presuming, therefore, that K648 is the source of the X-rays, it 
  would be both the first globular cluster and the most distant 
  X-ray detected PN.

Only four globular cluster PNe are known (see ABL and references
  therein), but their very existence implies an unusual
  evolutionary path.  
The lower the mass of a post-asymptotic giant branch (AGB) 
  remnant, the longer it takes to heat
  up to temperatures sufficiently hot to ionize
  the ejected envelope and produce a visible PN, by which time the
  ejecta may have dissipated. 
Recent HST studies of white dwarfs in globular clusters indicate that 
  AGB remnant masses are very low ($0.50
  \pm 0.02$ \msun; Renzini et al. 1996; Cool, Piotto \& King 1996;
  Richer et al. 1997). 
AGB remnants with these masses have evolutionary
  timescales that are much longer than PNe dissipation timescales
  (Sch\"{o}nberner 1983; Vassiliadis \& Wood 1994). 
Therefore, we should not expect to observe {\it any} PNe within 
  globular clusters! 
Jacoby et al. (1997) have suggested that the globular cluster PNe were
  therefore formed through close binary interactions. 

This certainly applies to the case of K648 since the most massive 
  main sequence stars in M15 should have masses $M \leq 0.8$~\msun, 
  with AGB remnant masses too low to
  produce PNe.  
Moreover, ABL measured an anomalously high
  mass of $0.60 \pm 0.02$~\msun\space for the central 
  star of K648.  
They suggest the most likely origin for this higher mass white
  dwarf is from mass augmentation of its progenitor star (i.e. 
  formation of  a blue straggler) due to either mass transfer in or 
  merger of a close binary. 
They also note that the morphology of K648 is typical of elliptical
  PNe.  
This could be accounted for by either nebular ejection during a common
  envelope phase, or by enhanced rotation of the AGB progenitor (spun 
  up in a binary system or during a stellar merger).  
In any event, a close binary evolutionary stage is indicated. 
Unfortunately, ABL were unable to find further evidence that the 
  central star is currently a close binary, as no
  variability was seen over their 7-day HST/WFPC2 observing campaign.

In principle the X-ray emission from K648 might arise from either diffuse
  thermal emission from a large cavity inside the nebula, or point-like
  emission from the PN nucleus (PNN; see Soker \& Kastner 2002 for a
  recent review). 
However, given its short-term variability (within 19 days), emission
  from the compact PNN is much more likely. 
Soker \& Kastner (2002) predict that 20--30\% of elliptical PNe
  probably have magnetically active late-type companions in binary
  systems with $a\siml 65$\rsun\space and X-ray luminosities of 
  $L\simg 10^{29}$~erg~s$^{-1}$.  
Indeed, the most likely scenario, in view of K648's X-ray variability, is
  some sort of binary system, where accretion either directly or
  indirectly is reponsible (J. Kastner, priv. comm.).

\section{Conclusions}

We have identified four faint X-ray sources in the globular cluster
  M15 using {\it Chandra}. 
Three of these are probable DNe (although the possibility 
that source B is a qSXT cannot be ruled out) and the fourth is
  associated with the planetary nebula K648.
One of the DNe (source A) was initially discovered by applying image
  subtraction to WFPC2 images of the core of M15 (Charles et
  al. 2001). 
In this study we reported on the detection of its X-ray emission.
The qSXT/DN (source B) and the other DN (source C) were first detected
  using {\sc wavdetect} in the {\it Chandra} images 
  and then associated with optical counterparts
  in HST images, which allowed us to speculate on their nature. 
We have also discovered that the PN K648 in M15 (source D) is a variable X-ray
  source, perhaps indicating a binary nature for the planetary nebula nucleus.
If K648 is indeed the source of the X-rays, it 
  is both the first globular cluster and the most distant 
  X-ray detected planetary nebula.

Sources A, B and C (both DNe and the qSXT/DN)
  had X-ray luminosities ranging from
  $2.34\times10^{31} - 3.66\times10^{32} \ {\rm erg} \ {\rm s}^{-1}$ in 
  the 0.5--7~keV range during these {\it Chandra} observations.
Converting the luminosities to the 0.5--2.5~keV range, one gets a 
  minimum luminosity of  $3.11\times10^{31} \ {\rm erg} \ {\rm
  s}^{-1}$ and a maximum of $2.27\times10^{32} \ {\rm erg} \ {\rm
  s}^{-1}$ -- these luminosities place these sources at the bright end
  of the galactic systems studied by Verbunt et al. (1997).
This implies that the majority of DNe in M15 remain to be
  discovered by more sensitive observations.
In order to identify optical counterparts of more {\it Chandra} sources, 
  we need deeper imaging of M15 (to identify very faint sources like qSXTs or
  magnetic CVs), as well as observations at different epochs (to catch
  quiescent DNe and X-ray transients in outburst). 

\section*{Acknowledgments}

DCH is grateful to the Academy of Finland and to PPARC for financial
support. MBD gratefully acknowledges the support of a Swedish Royal
Academy of Sciences (KVA) Research Fellowship.
The authors thank Craig Heinke, Bruce Balick and Joel
Kastner for valuable comments.  
The authors also wish to thank Jonathan C. McDowell for useful 
suggestions, Miriam Krauss at the {\it Chandra} HelpDesk,
and the anonymous referee for useful comments.
DCH is grateful to Panu Muhli for useful comments. 
This research has made use of NASA's Astrophysics Data
System, SAOImage DS9, developed by Smithsonian Astrophysical
Observatory, and of the SIMBAD database operated at CDS, Strasbourg, France.
Based on observations made with the NASA/ESA Hubble Space Telescope,
obtained from the Data Archive at the Space Telescope Science
Institute, which is operated by the Association of Universities for 
Research in Astronomy, Inc., under NASA contract NAS 5-26555.

\section*{References}

Alves D.R., Bond H.E., Livio M., 2000, AJ, 120, 2044 (ABL)

\noindent Anderson J., Cool A.M., King I.R., 2003, ApJ, 597, L137

\noindent Auri\`ere M., Le Fevre O., Terzan A., 1984, A\&A, 138, 415

\noindent Bailey J., 1980, MNRAS, 190, 119

\noindent Baumgardt H., Hut P., Makino J., McMillan S., Portegies Zwart S., 2003, ApJ, 582, L21

\noindent Bianchi L., Ford H., Bohlin R., de Marchi G., Paresce F.,
1995, A\&A, 301, 537

\noindent Bianchi L., Bohlin R., Catanzaro G., Ford H., Manchado A.,
2001, AJ, 122, 1538

\noindent Charles P.A., 2001, in Kaper L., van den Heuvel E.P.J, Woudt
P.A., eds, Black Holes in Binaries and Galactic Nuclei. Springer, p. 27

\noindent Charles P.A., Clarkson W.I., van Zyl L., 2002, NewA, 7, 21

\noindent Charles P.A., Jones D.C., Naylor T., 1986, Nature, 323, 417

\noindent Cool A.M., Piotto G., King I.R., 1996, ApJ, 468, 655

\noindent de Marchi G., Paresce F., 1994, ApJ, 422, 597

\noindent de Marchi G., Paresce F., 1996, ApJ, 467, 658

\noindent Di Stefano R., Rappaport S., 1994, ApJ, 423, 274

\noindent Djorgovski S., King I.R., 1986, ApJ, 305, L61

\noindent Dotani T. et al., 1990, Nature, 347, 534

\noindent Durrell P.R., Harris W.E., 1993, AJ, 105, 1420

\noindent Edmonds P.D., Heinke C.O., Grindlay J.E., Gilliland R.L.,
2002, ApJ, 564, L17

\noindent Edmonds P.D., Gilliland R.L., Heinke C.O., Grindlay J.E.,
2003, ApJ, 596, 1997

\noindent Ferraro F.R., Paresce F., 1993, AJ, 106, 154

\noindent Gebhardt K., Pryor C., Williams T.B., Hesser J.E., 1992, BAAS, 24, 1188

\noindent Gerssen J., van der Marel R.P., Gebhardt K., Guhathakurta P., Peterson R.C., Pryor C., 2002, AJ, 124, 3270

\noindent Gerssen J., van der Marel R.P., Gebhardt K., Guhathakurta P., Peterson R.C., Pryor C., 2003, AJ, 125, 376

\noindent Giacconi R., et al., 2001, ApJ, 551, 624

\noindent Grabhorn R.P., Cohn H.N., Lugger P.M., Murphy B.W., 1992, ApJ, 392, 86
\noindent Guerrero M.A., Chu Y.-H., Gruendl R.A., 2000, ApJS, 129, 295

\noindent Guerrero M.A., Chu Y.-H., Gruendl R.A., Williams R.M.,
Kaler J.B., 2001, ApJ, 553,  L55

\noindent Guhathakurta P., Yanny B., Schneider D.P., Bahcall J.N., 1996, AJ, 111, 267

\noindent Hakala, P.J., Charles, P.A., Johnston, H.M., Verbunt, F., 1997, 
    MNRAS, 285, 693

\noindent Harris W.E,. 1996, AJ, 112, 1487 

\noindent Heinke C.O., Edmonds P.D., Grindlay J.E., 2001, ApJ, 562, 363

\noindent Heinke C.O., Edmonds P.D., Grindlay J.E., Lloyd D.A., Cohn
H.N. \& Lugger P.M., 2003b, ApJ, 590, 809

\noindent Heinke C.O., Grindlay J.E., Lugger P.M., Cohn H.N.,
Edmonds P.D., Lloyd D.A. \& Cool A.M., 2003a, ApJ, 598, 501

\noindent Heinke C.O., et al., 2004, submitted to ApJ

\noindent Ho L.C., Terashima Y., Okajima  T., 2003, ApJ, 587, L35

\noindent Homer L., Anderson S.F., Margon B., Downes R.A., Deutsch
E.W., 2002, AJ, 123, 3255

\noindent Ilovaisky S.A., Auri\`{e}re M., Koch-Miramond L., Chevalier C., Cordoni J.-P., Crowe R.A., 1993, A\&A, 270, 139

\noindent Ioannou Z., Naylor T., Smale A.P., Charles P.A., Mukai K., 2002, A\&A, 382, 130

\noindent Jacoby G.H., et al., 1997, AJ, 114, 2611

\noindent Kaluzny J., Thompson I., Krzeminski W., Pych W., 1999,
A\&A, 350, 469

\noindent Kastner J.H., Soker N., Vrtilek S.D., Dgani R., 2000, ApJL, 545, 57

\noindent Kastner J.H., Vrtilek S.D., Soker N., 2001, ApJL, 550, 189

\noindent Kulkarni S.R., Anderson S.B., 1996, IAUS, 174, 181

\noindent Kuulkers E., Norton A., Schwope A., Warner B., 2003,
astrop-ph/0302351

\noindent Lauer T.R., et al. 1991, ApJ, 369, L45

\noindent McClintock J.E., Remillard R.A., 2000, ApJ, 531, 956

\noindent McNamara B.J., Harrison T.E., Baumgardt H., 2004, ApJ,
602, 264

\noindent Murray S.S. et al., 1997, Proc SPIE, 3114, 11

\noindent Newell B., O'Neill E.J., 1978, ApJS, 37, 27

\noindent Peterson R.C., Seitzer P., Cudworth K.M., 1989, ApJ, 347, 251

\noindent Phinney F.S. 1996, in Milone E.F., Mermilliod J.-C., eds,
ASP Conf. Ser. Vol. 90, The Origins, Evolution, and Destinies of
Binary Stars in Clusters. Astron. Soc. Pac., San Francisco, p.163

\noindent Pooley D., et al. 2002a, ApJ, 569, 405

\noindent Pooley D., et al. 2002b, ApJ, 573, 184

\noindent Renzini A., et al., 1996, ApJ, 465, L23

\noindent Richer H.B., et al.,  1997, ApJ, 484, 741

\noindent Rutledge R.E., Bildsten L., Brown E.F., Pavlov G.G.,
Zavlin V.E., 2001, ApJ, 559, 1054

\noindent Sch\"onberner D., 1983, ApJ, 272, 708

\noindent Simon V., 2002, A\&A, 381, 151

\noindent Soker N., Kastner J.H., 2002, ApJ, 570, 245

\noindent Sosin C., King I.R., 1997, AJ, 113, 1328

\noindent Swank J.H., Boldt E.A., Holt S.S., Rothschild R.E.,
Serlemitsos P.J., 1978, ApJ, 226, L133

\noindent Tuairisg S.\'O., Butler R.F., Shearer A., Redfern R.M., Butler
D. \& Penny A., 2003, MNRAS, 345, 960

\noindent van der Marel R.P., Gerssen J., Guhathakurta P., Peterson R.C., 
Gebhardt K., 2002, AJ, 124, 3255

\noindent van Zyl L., 2002, PhD thesis, Univ. Oxford

\noindent Vassiliadis E., Wood P.R., 1994, ApJS, 92, 125

\noindent Verbunt F., Bunk W.H., Ritter H., Pfeffermann E., 1997,
A\&A, 327, 602

\noindent Verbunt F., Lewin W.H.G., 2004, to appear in Lewin W.H.G., 
van der Klis M., eds, Compact Stellar X-ray Sources. Cambridge
University Press. astro-ph/0404136

\noindent Verbunt F., Wheatley P.J., Mattei J.A., 1999, A\&A,
346, 146

\noindent Warner B., 1995, Cataclysmic Variable Stars, Cambridge
Univ. Press, Cambridge

\noindent White N.E., Angelini L., 2001, ApJ, 561, L101

\noindent Wijnands R., Heinke C.O., Grindlay J.E., 2002, ApJ, 572, 1002

\end{document}